# The Acousto-Thermoelectric Effect

*Hasan Salehi Najafabadi[1*], Mark. A. Meier[1†], Gary A. Hallock[2‡]*

[1]Department of Physics, University of Houston, Houston, TX, 77204, USA
[2]Department of Electrical and Computer Engineering, University of Texas at Austin, Austin, TX 78712, USA

[*]Corresponding Author
[†]mameier@uh.edu

[*]hsalehinajafabadi@uh.edu
[‡]hallock@utexas.edu

## Abstract

An effect we have termed the acousto-thermoelectric effect is theorized for temperature gradients driven by acoustic modulation. The effect produces a dynamic and spatially varying voltage. Adiabatic acoustic fluctuations in a solid cause temperature variations and temperature gradients that generate quasi-static thermoelectric effects correlated with the time and spatial scales of the acoustic fluctuations. This phenomenon is distinctive from the static thermoelectric effect in that the hot spots (heat sources) and cold spots (heat sinks) change locations and vary over short time scales. Predictions are made for a semiconductor material, indium antimonide, showing that the effect is measurable under laboratory conditions. The sample is excited by a resonant acoustic mode with frequency 230 kHz, wavelength of 1.37 cm, and pressure amplitude of 2.23 MPa (rms). The predicted peak voltage between positions where maximum and minimum temperatures occur is 2.6 $\mu$V. The voltage fluctuates with the same frequency as acoustic resonance.

The thermoelectric effect has been widely studied for many different materials[1–15]. Many studies examine static thermal gradients created with a heat source located at one end of a thermoelectric sample and/or a heat sink located at the other end. We theorize that adiabatic pressure fluctuations from an acoustic wave drive dynamic thermal gradients and generate thermoelectric effects having temporal and spatial variations corresponding to the acoustic wave.

A pressure change in a local region of a lattice can be treated as adiabatic if heat flow is much slower than the pressure dynamics. An acoustic pressure change occurs rapidly allowing little time for heat to diffuse in the lattice. This causes a local temperature modulation related to the pressure modulation. Correspondingly, a rapid dynamic pressure gradient in the lattice produced by an acoustic wave leads to a dynamic temperature gradient. The temperature gradient produces a thermoelectric effect and electrical response. We name this newly discovered phenomenon the acousto-thermoelectric (ATE) effect.

This paper focuses on the conditions for, and characteristics of the ATE effect. First, adiabatic pressure modulation in a solid is discussed, followed by temperature modulation from adiabatic pressure change. We argue that the conditions generating thermoelectric effects may be treated as quasi-static even though the dynamics are rapid compared to thermal diffusion time scales. Then, the ATE effect that derives from temperature modulation is described. An example ATE voltage is determined between the antinodes of a standing acoustic wave, where maximum and minimum temperatures occur, in an indium antimonide (InSb) sample.

This study is not about photoacoustic effect or measuring thermal properties. Photoacoustic effect is when an acoustic signal is produced by illuminating a sample in an enclosed cell with periodically modulated (chopped) light. Also, this study is not about characterization of thermal properties of materials. This work describes the thermoelectric effect due to charge carrier transport in bulk of a solid caused by acoustically driven temperature gradient.

A thermodynamic process in which there is no heat exchange between a system and its surroundings is defined to be adiabatic. The heat gained or lost by the system is zero. This is an idealized concept; nonetheless, a system in which heat diffusion is slow compared to the dynamics of pressure change can be reasonably modelled as adiabatic. During an adiabatic process, the temperature of a system changes when pressure changes.

Suppose there is a sudden thermal fluctuation in a lattice. The heat equation, or thermal diffusion equation, is given by

$$\frac{\partial T}{\partial t} = D_{th} \nabla^2 T. \tag{1}$$

The coefficient $D_{th}$ in this equation is thermal diffusivity and $T$ is temperature. The heat equation may be evaluated for an initial temperature profile in one dimension given by $T(x, t = 0) = T_o \cos(kx) + T_A$, where position and time are given by $x$ and $t$, respectively, $k = 2\pi/\lambda$ is the wavenumber and $\lambda$ is wavelength, and $T_A$ is the ambient temperature. This profile is shown in Fig. 1.

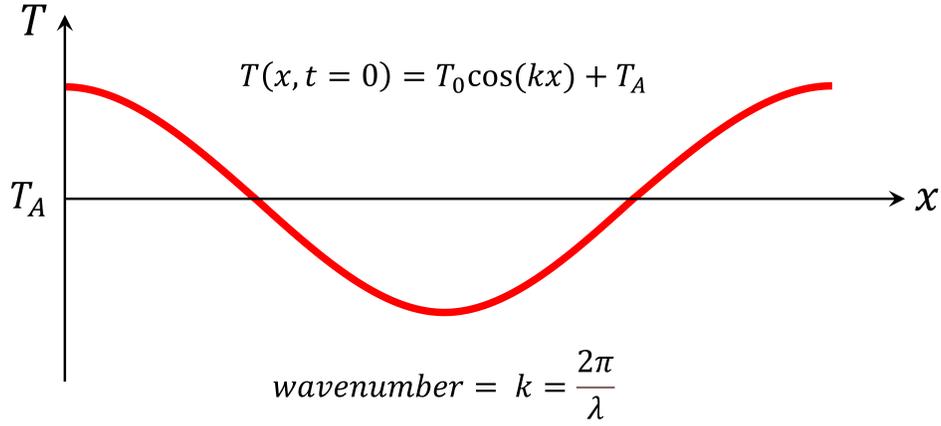

Fig. 1: An initial temperature profile in the lattice is shown. The wavenumber, $k$, relates to the spatial change of the temperature profile.

The solution is given by,

$$T(x,t) = T_0 \cos(kx)\, e^{-\frac{4\pi^2}{\lambda^2} D_{th} t} + T_A. \tag{2}$$

The exponential term may be written in the form $-t/\tau$, where $\tau$ is the characteristic time given by

$$\tau = \frac{\lambda^2}{4\pi^2 D_{th}}. \tag{3}$$

If the period of acoustic wave, given by $t_{aw}$, is much smaller than the characteristic time,

$$t_{aw} \ll \tau, \tag{4}$$

then the acoustic fluctuations happen much faster than heat diffusion, and very little heat is transferred during the pressure change. Therefore, this process may be treated as adiabatic.

From the thermoelastic effect and Kelvin's formula, the ratio of temperature change to pressure change in an adiabatic process can be represented for a solid. Replacing stress change in the thermoelastic stress analysis relationship for a solid given by Pitarresi[16] by pressure change gives a relationship of the same form,

$$\left(\frac{\Delta T}{\Delta P}\right)_a = \frac{\kappa_T \alpha}{C} T, \tag{5}$$

where $\Delta T$ and $\Delta P$ are the amplitude of temperature change and pressure change, respectively, $\kappa_T$ is the thermal expansion coefficient, $\alpha$ is specific volume, and $C$ is specific heat. This implies that the temperature modulation is proportional to the pressure modulation, $\Delta T = \gamma \Delta P$, where $\gamma$ is the proportionality constant given by the right hand side of equation (5). Adiabatic pressure changes can be used to produce a dynamic temperature wave in a solid by means of an acoustic wave. Acoustic compression in the lattice causes lattice heating and decompression causes lattice cooling. Therefore hot regions and cold regions are produced, as well as thermal gradients in between, that change over time.

Mobile charge carriers in a lattice respond quickly to acoustic perturbations. We approximate the time scale of response to the period of the plasma frequency. Plasma frequency is given by

$$\omega_p = \left(\frac{n_0 e^2}{\varepsilon m^*}\right)^{1/2}, \tag{6}$$

where $n_o$ is the number density of charge carriers, e is elementary charge, $\varepsilon$ is permittivity, and $m^*$ is the electron's effective mass. The corresponding period is given by

$$t_P = \frac{2\pi}{\omega_p}. \tag{7}$$

If the time scale of electron transport, $t_p$, is much shorter than the time scale of a perturbation, then the electrical response of the system to the perturbation can be treated as though the system is static, even though the system is evolving relatively slowly. This quasi-static assumption is used in considering the thermoelectric response to a temperature gradient evolving on the time scale of acoustical dynamics.

In metals, electron densities are high, and the plasma period may typically be on the order of a femtosecond. In semiconductors, plasma periods may typically range on orders near a picosecond. Therefore, for acoustical phenomena having periods from microseconds to milliseconds, conditions for a quasi-static assumption often hold.

An example is shown for the case described. A room temperature sample of InSb has an intrinsic number density of $2.05 \times 10^{22}$ m$^{-3}$,[17] a permittivity of 16.8 times the permittivity of free space ($8.85 \times 10^{-12}$ Fm$^{-1}$),[18] and an effective mass of 0.014 times the mass of an electron ($9.11 \times 10^{-31}$ kg).[18] Given the elementary charge of $1.60 \times 10^{-19}$ C, the plasma frequency is computed to be

$$\omega_p = 1.7 \times 10^{13} \, \frac{\text{rad}}{\text{s}}, \tag{8}$$

and the corresponding period is

$$t_P = 3.8 \times 10^{-13} \text{ s}. \tag{9}$$

The acoustic resonance frequency is 230 kHz, giving an acoustic period of $t_{aw} = 4.3 \times 10^{-6}$ s. The ratio of the plasma frequency period and acoustic resonance period is

$$\frac{t_P}{t_{a\omega}} = 8.7 \times 10^{-8}. \tag{10}$$

Given the much shorter response time of electrons relative to the acoustical perturbation, the quasi-static condition is met, allowing a static treatment of the thermoelectric response.

Adiabatic pressure modulation in a solid produces larger temperature modulation if the ratio **ΔT/ΔP** given in equation (5) has a larger value. In this case, a larger temperature gradient results and, if the material is thermoelectric, the corresponding thermoelectric effect is also larger. Therefore, a thermoelectric material possessing the characteristics of larger specific volume and thermal expansion coefficient, and smaller specific heat is favorable to the ATE effect.

Figure 2 shows a standing acoustic wave at the time of peak amplitude, $t = t_{peak}$, and one-half period later, $t = t_{peak} + t_{aw}/2$. At time $t_{peak}$, there is a hot region corresponding with high pressure and a cold region corresponding with low pressure. If quasi-static conditions are met, we may evaluate the thermoelectric effect at time $t_{peak}$ as if the temperature profile is static. For example, the Seebeck effect[19] is illustrated for the case of a negative Seebeck coefficient. In this case, the cold region develops a voltage greater than the hot region. One half-period later, the hot region changes into a cold region, and the cold region changes into a hot region. Thus, the ATE voltage changes polarity every half-period, and an alternating voltage is produced across the lattice. In between the times of peak amplitudes, the pressure and temperature amplitudes subside and so do the voltage differences, even reaching zero as the pressure and temperature profiles become flat.

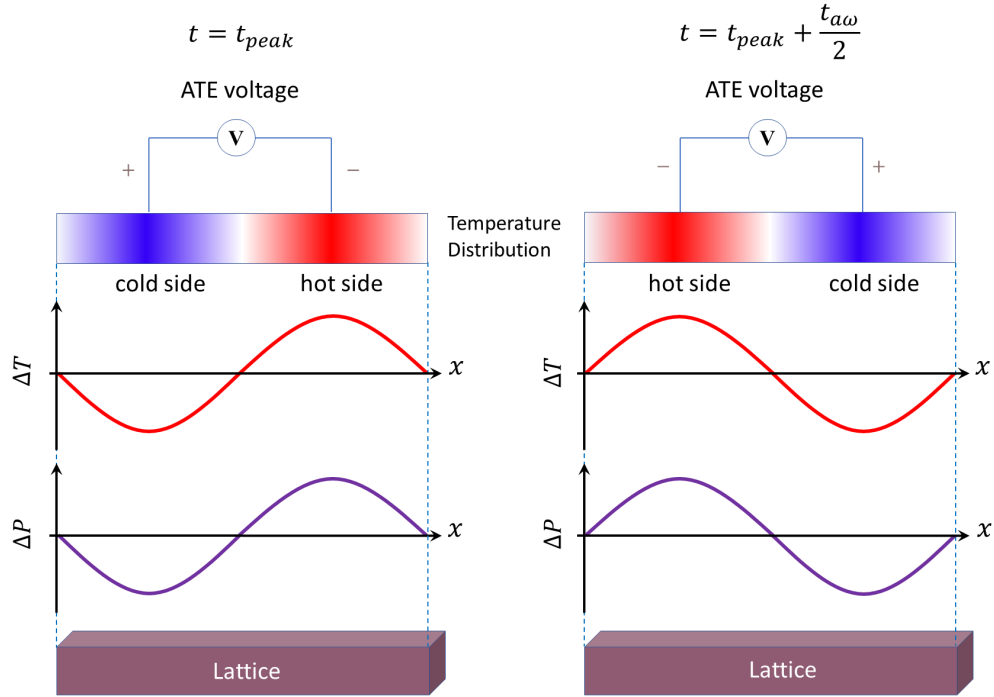

Fig. 2: The acousto-thermoelectric (ATE) effect is illustrated. Profiles related to the times of peak amplitudes, one-half period apart, are shown. The blue and red lines represent adiabatic pressure modulations and corresponding temperature modulations, respectively. The color scale is a schematic representation of temperature distribution. The ATE voltage alternates between half-periods.

An important distinguishing characteristic exists between the ATE effect and the conventional thermoelectric effect, particularly regarding the role of thermal conductivity. The conventional thermoelectric effect is a static phenomenon, meaning the hot side and cold side do not change, and a static voltage is created between the hot and cold locations. Materials that have high electrical conductivity and low thermal conductivity are desired for the best results. These factors translate into a fundamental dimensionless parameter referred to as the thermoelectric figure of merit, $ZT^6$. Using the Wiedemann–Franz law, it can be shown that these factors tend to contradict one another and limit the achievement of high thermoelectric efficiency[6,20,21]. However, the objective for low thermal conductivity may be considerably relaxed for the ATE effect. This is because the time scales of acoustic fluctuations are considerably shorter than the effective time scale of the static case. The thermal conductivity need only be sufficiently low so that relatively little heat transport occurs within the period of acoustic fluctuation. Therefore, many materials that may not qualify for high thermoelectric efficiency may, nonetheless, qualify as good materials for the ATE effect.

This section evaluates the theorized ATE voltage between points of maximum and minimum temperatures created by a resonant acoustic wave in an InSb crystal. Acoustic resonance of the crystal has been reported in our previous publications[17,22–28]. We use the pressure amplitudes produced in our previous experimental work[17], and the Seebeck formula for this evaluation. Fig. 3 shows a schematic of the InSb crystal attached to a piezoelectric (PZT) crystal. We refer to this configuration as the InSb transducer. The PZT drives a standing acoustic wave in the InSb crystal with an amplitude of 2.23 MPa (rms) and a frequency of 230 kHz.

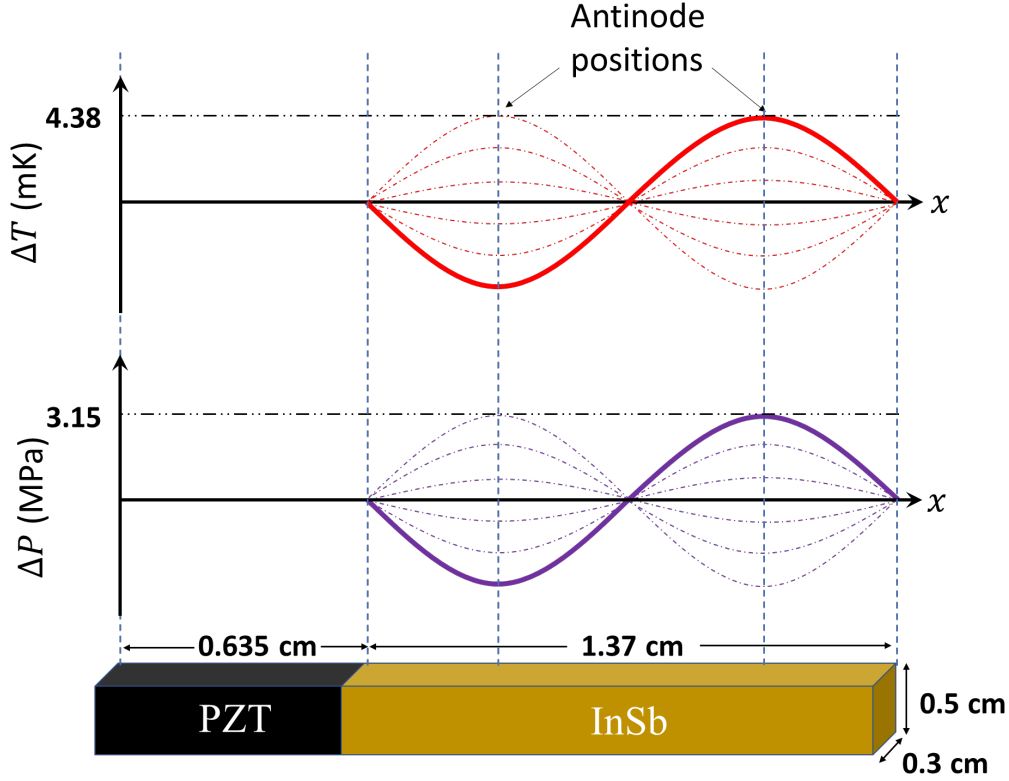

Fig. 3. Schematic representation of pressure modulation and corresponding temperature modulation in the InSb transducer. Purple and red lines show the pressure modulation ($\Delta P$) and acoustically driven temperature modulation ($\Delta T$), respectively, as a function of position.

The thermal diffusivity of InSb is[18]

$$D_{th} = 0.16\ \frac{\text{cm}^2}{\text{s}}. \qquad (11)$$

The frequency of the acoustic wave for $(3/2)\lambda$ resonance mode in the InSb transducer is 230 kHz, and the related wavelength and period is $\lambda = 1.37$ cm, and $t_{aw} = 4.3 \times 10^{-6}$ s, respectively. Plugging these values into equation (3) gives the thermal diffusion characteristic time,

$$\tau = 0.30 \text{ s}. \tag{12}$$

Thus,

$$t_{aw} \ll \tau, \tag{13}$$

and we treat the pressure modulation as adiabatic.

Since very little heat transfer takes place within a period of the acoustic modulation, the lattice temperature changes locally in accordance with the pressure change. The temperature modulation due to adiabatic pressure modulation for InSb is obtained using equation (5). Plugging in the values $(5770 \text{ kg/m}^{-3})^{-1}$ for specific volume,[18] $5.37 \times 10^{-6} \text{ K}^{-1}$ for thermal expansion coefficient,[18] 200 J/(kg K) for specific heat,[18] and 300 K for room temperature gives

$$\left(\frac{\Delta T}{\Delta P}\right)_{a,InSb} = 1.4 \times 10^{-9} \frac{\text{K}}{\text{Pa}}. \tag{14}$$

The peak amplitude of pressure modulation is the square root of two times the rms amplitude, therefore $\Delta P_{max} = 3.15$ MPa. Then, applying equation (14), the corresponding peak amplitude of temperature modulation is given by

$$\Delta T_{max} = 4.4 \text{ mK}. \tag{15}$$

Hence, the difference between the hot and cold side temperature, or peak-to-peak temperature amplitude, is

$$\Delta T = T_{hot} - T_{cold} = 8.8 \text{ mK} \tag{16}$$

The Seebeck coefficient for intrinsic InSb at room temperature is around $S$= -300 µV/K[29]. Therefore, the Seebeck formula gives a corresponding voltage magnitude of approximately

$$|\Delta V| = |S\Delta T| = 2.6\ \mu\text{V}. \qquad (17)$$

Fig. 4. Shows the ATE voltage between positions of temperature antinodes, as a function of time, for a full period of acoustic resonance in the InSb transducer shown in Fig. 3.

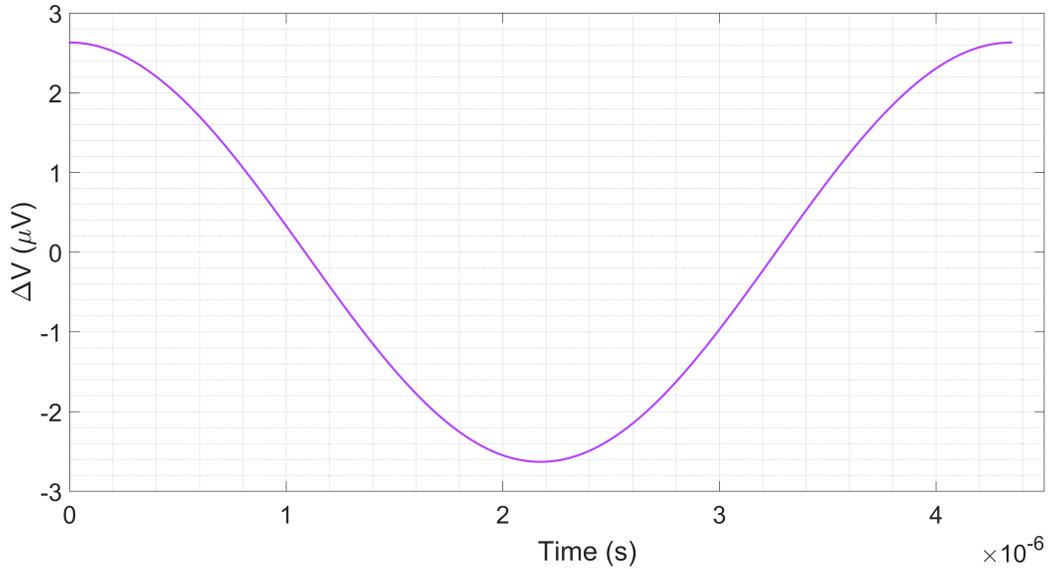

Fig. 4. ATE voltage between positions of temperature antinodes as a function of time for a full time period of acoustic resonance.

The acousto-thermoelectric (ATE) effect is predicted for acoustic waves in a thermoelectric material. A characteristic time for heat transport is obtained from the heat equation, and depends on acoustic wavelength and thermal diffusivity. If the period of pressure modulation from the acoustic wave is very much shorter than the characteristic time for heat diffusion, then there is very little heat transport and the process is treated as adiabatic. According to the thermodynamics of solids, the temperature modulation due to an adiabatic pressure fluctuation in the lattice is proportional to the specific volume, thermal expansion coefficient, pressure modulation, and temperature of the lattice and is inversely proportion to the specific heat of the lattice.

When acoustic fluctuations are adiabatic, a compression in the lattice causes a temperature increase, and a decompression causes a temperature decrease. Therefore, there are cold and hot regions associated with acoustic fluctuations, and these regions are changing with the changes in acoustic pressure. For time harmonic acoustic waves, the cold side changes to a hot side and the hot side changes to a cold side over a time interval equal to half the period. Thus, a dynamic acoustically-driven thermoelectric effect exists and produces an alternating voltage across the lattice. The ATE voltage fluctuates with the same frequency as the acoustic wave.

The approximate amplitude of ATE voltage between positions of maximum and minimum temperatures is computed for a known acoustic resonance in an InSb transducer. The mode frequency is 230 kHz, the wavelength is 1.37 cm, and the pressure amplitude is 2.23 MPa (rms). The ATE voltage is predicted to be about 2.6 $\mu$V and fluctuates with the same frequency as the mode frequency.

We believe the ATE effect may be considered more broadly for acoustic waves in elastocaloric or thermoelastic materials depending on thermoelectric properties. Adiabatic compression and decompression from an acoustic wave change the lattice temperature and introduces a temperature gradient and a possible corresponding ATE effect. If the ratio *ΔT*/*ΔP* is large for some elastocaloric materials, then the temperature gradients and ATE voltages may be increased. This effect might be useful for sensing pressure gradient, since pressure gradient causes a temperature gradient which results in an electrical response. The effect might also be seen in other structures like thin films, quantum wells or rods, and some 2D materials. Applications might include sensors, transducers.

## Acknowledgements

This work was funded by the University of Houston.

## Conflict of Interest

The authors have no conflicts to disclose.

## Data availability

The data that support the findings of this study are available from the corresponding author upon reasonable request.